\documentstyle[11pt]{article}
 
\begin{document}

\title{The Optical Gravitational Lensing Experiment.\\ 
The Catalog of Periodic Variable Stars in the Galactic Bulge.\\ 
V.~Periodic Variables in Fields:\\
MM5-A, MM5-B, MM7-A and MM7-B\footnote{Based on observations obtained at the 
Las Campanas Observatory of the Carnegie Institution of Washington}} 

\author{A.~~U~d~a~l~s~k~i$^{1}$,
~A.~~O~l~e~c~h$^1$,
~M.~~S~z~y~m~a~\'n~s~k~i$^{1}$,
~J.~~K~a~\l~u~\.z~n~y$^{1}$,\\
~M.~~K~u~b~i~a~k$^{1}$,
~M.~~M~a~t~e~o$^2$,
~W.~~K~r~z~e~m~i~\'n~s~k~i$^3$~~and~~K.~Z.~~S~t~a~n~e~k$^{4}$}
\date{$^1$Warsaw University Observatory, Al.~Ujazdowskie~4, 00--478~Warszawa, Poland\\
e-mail: (udalski,olech,msz,jka,mk)@sirius.astrouw.edu.pl\\
$^2$Department of Astronomy, University of Michigan, 821~Dennison Bldg., Ann Arbor, MI~48109--1090, USA\\
e-mail: mateo@astro.lsa.umich.edu\\
$^3$Carnegie Observatories, Las Campanas Observatory, Casilla~601, La~Serena, Chile\\
e-mail: wojtek@roses.ctio.noao.edu\\
$^4$Harvard-Smithsonian Center for Astrophysics, 60~Garden Street, MS~20, Cambridge, MA~02138\\
e-mail: kstanek@cfa.harvard.edu}

\maketitle

\abstract{We present the fifth part of the OGLE Catalog of Periodic Variable 
Stars in the Galactic bulge. 574 variable stars: 44 pulsating, 369 eclipsing 
and 161 miscellaneous type were detected in four fields located symmetrically 
in galactic lattitude around the Galactic center: MM5-A, MM5-B, MM7-A and 
MM7-B. 

The Catalog and individual observations are available in digital form
from the OGLE Internet archive.}

\noindent {\bf Key words:} Catalogs -- Stars: variables

\section{Introduction}

The huge databases of photometric observations collected during microlensing 
searches provide an unique material for studying stellar variability. Data 
collected during the Optical Gravitational Lensing Experiment (OGLE) -- long 
term observing project with the main goal of searching for  dark matter in our 
Galaxy with microlensing phenomena (Paczy\'nski 1986, Udalski et al. 1994a) -- 
have already been  searched for variable stars and four parts of the OGLE 
Catalog of Periodic Variable Stars have been published (Udalski et al. 1994b, 
1995a, 1995b, 1996 hereafter referred as Paper I--IV). The Catalog includes 
2287 periodic objects from the Baade's Window in the Galactic bulge. Paper III 
presents also simple analysis of the completeness of the Catalog. 

This paper is a continuation of the Catalog and presents periodic variable 
stars found in four fields located symmetrically in galactic longitude around 
the Galactic center. 

\section{The Catalog}

The photometric data presented here were collected during four observing 
seasons of the OGLE microlensing search starting from April 13, 1992 through  
August 19, 1995. Observations were made at Las Campanas Observatory, Chile 
which is operated by Carnegie Institution of Washington. 1-m Swope telescope 
equipped with ${2048\times 2048}$ Ford/Loral CCD detector was used. Details of 
data pipeline, reduction technique and period search technique can be found in 
Paper~I. 

Present edition of the Catalog contains periodic variable stars with $\langle 
I\rangle$ brighter than 18~mag. In the following updates the Catalog will be 
extended toward fainter stars. There is also a lower limit of magnitude: 
${I\approx14}$ -- resulting from saturation of stellar images on CCD frames. 
The period search was limited to periods within 0.1--100 days range. 

This  part of the Catalog presents variable stars from four fields: MM5-A, 
MM5-B, MM7-A and MM7-B. Each of the fields covers approximately 
${15' \times 15'}$ on the sky. Equatorial and galactic coordinates of 
these fields are given in Table~1. Fields MM5-A and MM5-B as well as MM7-A and 
MM7-B overlap by about $1'$. Variable objects detected in the overlapping 
regions appear in the Catalog only once. 
\vspace{7pt}
\begin{center}
Table 1\\
Coordinates of MM5-A, MM5-B, MM7-A and MM7-B fields.\\
\vspace{7pt}

\begin{tabular}{|l|c|c|c|c|}
\hline
Field& $\alpha_{2000}$ & $\delta_{2000}$ & $l$ & $b$\\
\hline
MM5-A & 17$^h$47$^m$30$^s$ & --34$^\circ$45'00'' & --4.8 & --3.4\\
MM5-B & 17$^h$47$^m$30$^s$ & --34$^\circ$57'00'' & --4.9 & --3.5\\
MM7-A & 18$^h$10$^m$53$^s$ & --25$^\circ$54'20'' & ~~5.4 & --3.3\\
MM7-B & 18$^h$11$^m$47$^s$ & --25$^\circ$54'20'' & ~~5.5 & --3.5\\
\hline
\end{tabular}
\end{center}
\vspace{7pt}

The structure of the Catalog is identical as in the previous parts 
(Papers I--IV). Detected variable stars from each field are grouped into three 
categories: pulsating stars, eclipsing stars and miscellaneous type variables. 
The latter category consists of stars which cannot be classified unambiguously 
as pulsating or eclipsing stars. It contains mostly late type, 
chromospherically active stars, and likely some ellipsoidal variables. 

For every field and group of stars the Catalog consists of a table with basic 
parameters for every periodic variable object and an atlas containing the 
phased light curves and ${30''\times30''}$ finding charts -- part of the 
$I$-band frames. North is up and East is to the left on these charts. 

The basic parameters for every object include star designation, right 
ascention and declination (J2000), period in days and heliocentric Julian Date  
of maximum light (minimum for eclipsing variables), $I$ magnitude at maximum 
brightness, ${V-I}$ color at maximum brightness, $I$-band amplitude, 
classification and eventual remarks. 

Designation of the object follows the scheme introduced in Paper~I: OGLE {\it 
field} V{\it number}, e.g. OGLE MM5-A V22. The variable stars in every field 
are initially sorted according to magnitude. Thus lower number means brighter 
star. 

The equatorial coordinates of variable stars were calculated using 
transformation derived from position of stars from the HST Guide Star Catalog 
(Lasker et al. 1988). Typically about 15--20 GSC stars were identified in each 
field. Accuracy of coordinates is about 1~arcsec. 

Because of strategy adopted in the microlensing search, the vast majority of 
measurements was obtained in the $I$-band (typically 90--150 observations). 
About 30 or less $V$-band measurements were collected for each field during 
the entire search for color information. 

Classification within pulsating and eclipsing star groups follows the scheme 
of the General Catalog of Variable Stars (Kholopov et al. 1985). 

\section{Catalog of Periodic Variable Stars of the MM5-A, MM5-B, MM7-A and 
MM7-B Fields} 

Tables~2--13 and Appendices~A--L contain the catalog of pulsating, eclipsing 
and miscellaneous periodic stars for fields MM5-A, MM5-B, MM7-A and MM7-B, 
respectively. 

44 variable stars were classified as pulsating. Most of them are RR~Lyr stars 
type $ab$ and $c$. Remaining objects are short period $\delta$~Sct type 
pulsating stars. The periods of some of them fall below the lower limit (0.1 
day) of period search, but these stars were identified with ${2\times P}$ 
period. Thus they probably do not represent complete sample of this type of 
stars. 

369 eclipsing stars were identified in MM fields. The vast majority (277) of 
eclipsing objects belong to W~UMa type (EW). 60 Algol-type (EA) and 13 
$\beta$~Lyr-type (EB) stars were also identified. Many of Algol-type objects 
are detached systems ideal for precise determination of their parameters and 
distances with follow-up spectroscopic observations. 19 objects were 
classified as eclipsing (E) when the type of eclipse could not be determined  
unambiguously. 

The group of miscellaneous stars contains 161 objects. Most of them are red 
giants and subgiants, probably chromospherically active stars. Some objects in 
this group might be ellipsoidal variables what is indicated in the "Remarks" 
column. In such a case the period should be twice of that given in the Table. 

\vspace*{12pt}

\section{Summary}
We present the fifth part of the OGLE Catalog of Periodic Variable Stars in 
the Galactic bulge -- periodic variable stars from MM5-A, MM5-B, MM7-A and 
MM7-B fields. 574 periodic stars  were detected: 44 pulsating, 369 eclipsing 
and 161 miscellaneous type stars increasing the total number of periodic 
variable objects in the Catalog to 2861. 

The Catalog is supposed to be an open publication and regular updates are 
expected when more data become available and search for variables among fainter 
objects will be completed. Some errors, unavoidable in this first release of 
the Catalog, will also be corrected. Therefore we expect a feedback from the 
astronomical community when any errors, misclassification {\it etc.} are found. 

The Catalog and all individual observations of cataloged variable stars in 
both $V$ and $I$ bands (JD hel., magnitude, error) are available to the 
astronomical community from the OGLE Internet archive using anonymous ftp 
service from sirius.astrouw.edu.pl host (148.81.8.1), directory  
{\it /ogle/var\_catalog}. See README file in this directory. 

\vspace*{12pt}

\noindent {\bf Acknowledgments} It is a great pleasure to thank B. Paczy\'nski for valuable 
suggestions and discussions. This project was supported with the Polish KBN 
grants 2P03D02908 to A. Udalski and 2P03D02508 to M. Szyma\'nski
and the NSF grant AST-9530478 to B. Paczy{\'n}ski.

{\scriptsize
\vspace{7pt}
\begin{center}
Table 2\\
Pulsating Variable Stars in the MM5-A field\\
\vspace{7pt}

\begin{tabular}{lccrccccl}
Star ID  & $\alpha_{2000}$            & $\delta_{2000}$                &\multicolumn{1}{c}{$P$}       &T$_0$ --& $I$    & $(V{-}I)$ & $\Delta I$ & Type \\
OGLE      &                            &                                &           & 2~448~000 &        &        &       &               \\
MM5-A V6   &  17$^h$47$^m$19.$^s$08 &   --34$^\circ$52'45.''6 &   0.$^d$51202  &   724.0593 &  15.$^m$10 &   1.$^m$05 &  0.$^m$69 &   RRab \\
MM5-A V9   &  17$^h$47$^m$21.$^s$15 &   --34$^\circ$40'20.''8 &   0.$^d$51619  &   724.5310 &  15.$^m$27 &   1.$^m$25 &  0.$^m$72 &   RRab \\
MM5-A V15  &  17$^h$47$^m$50.$^s$53 &   --34$^\circ$49'25.''7 &   0.$^d$60022  &   724.3209 &  15.$^m$49 &   1.$^m$21 &  0.$^m$49 &   RRab \\
MM5-A V18  &  17$^h$47$^m$00.$^s$83 &   --34$^\circ$50'02.''0 &   0.$^d$52667  &   724.2808 &  15.$^m$54 &   1.$^m$31 &  0.$^m$74 &   RRab \\
MM5-A V19  &  17$^h$47$^m$33.$^s$22 &   --34$^\circ$39'20.''4 &   0.$^d$61251  &   724.2349 &  15.$^m$55 &   1.$^m$36 &  0.$^m$66 &   RRab \\
MM5-A V20  &  17$^h$46$^m$49.$^s$25 &   --34$^\circ$52'02.''2 &   0.$^d$39119  &   724.5387 &  15.$^m$58 &   1.$^m$08 &  0.$^m$25 &    RRc \\
\end{tabular}
\end{center}

\vspace{7pt}
\begin{center}
Table 3\\
Pulsating Variable Stars in the MM5-B field\\
\vspace{7pt}

\begin{tabular}{lccrccccl}
Star ID  & $\alpha_{2000}$            & $\delta_{2000}$                &\multicolumn{1}{c}{$P$}       &T$_0$ --& $I$    & $(V{-}I)$ & $\Delta I$ & Type \\
OGLE      &                            &                                &           & 2~448~000 &        &        &       &               \\
MM5-B V4   &  17$^h$47$^m$22.$^s$44 &   --34$^\circ$56'42.''2 &   0.$^d$52489  &   724.6038 &  14.$^m$89 &   1.$^m$08 &  0.$^m$56 &   RRab \\
MM5-B V10  &  17$^h$46$^m$48.$^s$22 &   --34$^\circ$51'33.''2 &   0.$^d$52391  &   724.4030 &  15.$^m$40 &   1.$^m$02 &  0.$^m$72 &   RRab \\
MM5-B V20  &  17$^h$47$^m$14.$^s$69 &   --34$^\circ$54'37.''6 &   0.$^d$64994  &   724.4972 &  15.$^m$66 &   1.$^m$15 &  0.$^m$46 &   RRab \\
MM5-B V21  &  17$^h$46$^m$51.$^s$35 &   --34$^\circ$56'44.''6 &   0.$^d$50359  &   724.3261 &  15.$^m$69 &        --      &  0.$^m$71 &   RRab \\
MM5-B V28  &  17$^h$46$^m$55.$^s$14 &   --34$^\circ$59'55.''1 &   0.$^d$47995  &   724.4437 &  15.$^m$88 &   1.$^m$06 &  0.$^m$62 &   RRab \\
MM5-B V31  &  17$^h$47$^m$29.$^s$18 &   --34$^\circ$59'33.''4 &   0.$^d$36456  &   724.5393 &  15.$^m$97 &   1.$^m$11 &  0.$^m$83 &   RRab \\
MM5-B V41  &  17$^h$47$^m$51.$^s$30 &   --35$^\circ$02'06.''3 &   0.$^d$25745  &   724.9370 &  16.$^m$15 &   0.$^m$94 &  0.$^m$30 &    RRc \\
MM5-B V44  &  17$^h$47$^m$46.$^s$43 &   --34$^\circ$52'57.''4 &   0.$^d$52311  &   724.2479 &  16.$^m$22 &   1.$^m$13 &  0.$^m$75 &   RRab \\
MM5-B V60  &  17$^h$47$^m$09.$^s$79 &   --34$^\circ$56'45.''2 &   0.$^d$126671 &   724.8166 &  16.$^m$56 &   1.$^m$02 &  0.$^m$34 &   DSCT \\
MM5-B V141 &  17$^h$47$^m$22.$^s$68 &   --34$^\circ$58'15.''4 &   0.$^d$180212 &   724.9432 &  17.$^m$82 &   1.$^m$36 &  0.$^m$24 &   DSCT \\
\end{tabular}
\end{center}

\vspace{7pt}
\begin{center}
Table 4\\
Pulsating Variable Stars in the MM7-A field\\
\vspace{7pt}

\begin{tabular}{lccrccccl}
Star ID  & $\alpha_{2000}$            & $\delta_{2000}$                &\multicolumn{1}{c}{$P$}       &T$_0$ --& $I$    & $(V{-}I)$ & $\Delta I$ & Type \\
OGLE      &                            &                                &           & 2~448~000 &        &        &       &               \\
MM7-A V2   &  18$^h$11$^m$00.$^s$10 &   --25$^\circ$57'38.''7 &   0.$^d$54078  &   724.3705 &  14.$^m$76 &   1.$^m$05 &  0.$^m$69 &   RRab \\
MM7-A V16  &  18$^h$11$^m$14.$^s$74 &   --25$^\circ$51'48.''4 &   0.$^d$56710  &   724.3438 &  15.$^m$48 &   1.$^m$12 &  0.$^m$74 &   RRab \\
MM7-A V20  &  18$^h$11$^m$18.$^s$69 &   --26$^\circ$00'42.''6 &   0.$^d$52984  &   724.3072 &  15.$^m$53 &   1.$^m$33 &  0.$^m$83 &   RRab \\
MM7-A V25  &  18$^h$10$^m$32.$^s$01 &   --25$^\circ$59'37.''3 &   0.$^d$49380  &   724.6095 &  15.$^m$56 &   2.$^m$49 &  0.$^m$66 &   RRab \\
MM7-A V26  &  18$^h$10$^m$48.$^s$98 &   --25$^\circ$56'30.''4 &   0.$^d$44361  &   724.6831 &  15.$^m$61 &   1.$^m$14 &  0.$^m$88 &   RRab \\
MM7-A V34  &  18$^h$11$^m$12.$^s$64 &   --25$^\circ$49'38.''7 &   0.$^d$53049  &   724.5836 &  15.$^m$76 &   1.$^m$18 &  0.$^m$57 &   RRab \\
MM7-A V44  &  18$^h$10$^m$54.$^s$53 &   --25$^\circ$49'42.''4 &   0.$^d$22759  &   724.8193 &  16.$^m$00 &   0.$^m$94 &  0.$^m$26 &    RRc \\
MM7-A V47  &  18$^h$10$^m$35.$^s$63 &   --25$^\circ$52'52.''4 &   0.$^d$28782  &   724.6755 &  16.$^m$04 &   1.$^m$24 &  0.$^m$29 &    RRc \\
MM7-A V88  &  18$^h$10$^m$48.$^s$71 &   --25$^\circ$52'58.''8 &   0.$^d$15886  &   724.7994 &  16.$^m$87 &   1.$^m$18 &  0.$^m$41 &   DSCT \\
MM7-A V94  &  18$^h$10$^m$49.$^s$76 &   --25$^\circ$48'23.''7 &   0.$^d$27499  &   724.7304 &  16.$^m$91 &   1.$^m$30 &  0.$^m$25 &    RRc \\
MM7-A V103 &  18$^h$10$^m$28.$^s$67 &   --25$^\circ$49'11.''3 &   0.$^d$078847 &   724.9533 &  17.$^m$06 &   1.$^m$12 &  0.$^m$11 &   DSCT \\
\end{tabular}
\end{center}

\pagebreak

\vspace{7pt}
\begin{center}
Table 5\\
Pulsating Variable Stars in the MM7-B field\\
\vspace{7pt}

\begin{tabular}{lccrccccl}
Star ID  & $\alpha_{2000}$            & $\delta_{2000}$                &\multicolumn{1}{c}{$P$}       &T$_0$ --& $I$    & $(V{-}I)$ & $\Delta I$ & Type \\
OGLE      &                            &                                &           & 2~448~000 &        &        &       &               \\
MM7-B V6   &  18$^h$12$^m$03.$^s$45 &   --25$^\circ$58'22.''9 &   0.$^d$53913  &   724.5066 &  15.$^m$32 &   1.$^m$15 &  0.$^m$66 &   RRab \\
MM7-B V7   &  18$^h$11$^m$51.$^s$16 &   --25$^\circ$58'12.''1 &   0.$^d$59002  &   724.4478 &  15.$^m$40 &   1.$^m$20 &  0.$^m$62 &   RRab \\
MM7-B V8   &  18$^h$12$^m$00.$^s$30 &   --25$^\circ$53'20.''2 &   0.$^d$55940  &   724.5959 &  15.$^m$43 &   1.$^m$20 &  0.$^m$51 &   RRab \\
MM7-B V12  &  18$^h$11$^m$36.$^s$75 &   --25$^\circ$49'07.''0 &   0.$^d$88121  &   723.7237 &  15.$^m$58 &   1.$^m$45 &  0.$^m$26 &   RRab \\
MM7-B V13  &  18$^h$11$^m$40.$^s$12 &   --25$^\circ$51'03.''1 &   0.$^d$53321  &   724.1109 &  15.$^m$59 &   1.$^m$40 &  0.$^m$52 &   RRab \\
MM7-B V14  &  18$^h$11$^m$33.$^s$48 &   --25$^\circ$55'08.''6 &   0.$^d$28675  &   724.8570 &  15.$^m$65 &   1.$^m$11 &  0.$^m$25 &    RRc \\
MM7-B V63  &  18$^h$11$^m$47.$^s$91 &   --25$^\circ$52'23.''5 &   0.$^d$61806  &   724.7452 &  16.$^m$99 &   1.$^m$37 &  0.$^m$74 &   RRab \\
MM7-B V87  &  18$^h$11$^m$55.$^s$38 &   --25$^\circ$55'19.''7 &   0.$^d$065833 &   725.0135 &  17.$^m$48 &   1.$^m$03 &  0.$^m$04 &   DSCT \\
\end{tabular}
\end{center}

\vspace{7pt}
\begin{center}
Table 6\\
Eclipsing Variable Stars in the MM5-A field\\
\vspace{7pt}

\begin{tabular}{lccrccccl}
Star ID  & $\alpha_{2000}$            & $\delta_{2000}$                &\multicolumn{1}{c}{$P$}       &T$_0$ --& $I$    & $(V{-}I)$ & $\Delta I$ & Type \\
OGLE      &                            &                                &           & 2~448~000 &        &        &       &               \\
MM5-A V3   &  17$^h$47$^m$26.$^s$92 &   --34$^\circ$43'16.''8 &   0.$^d$64733  &   724.1143 &  14.$^m$80 &   0.$^m$88 &  0.$^m$86 &     EW \\
MM5-A V7   &  17$^h$47$^m$14.$^s$16 &   --34$^\circ$38'39.''9 &  12.$^d$58860  &   713.4463 &  15.$^m$18 &        --      &  0.$^m$20 &     E? \\
MM5-A V13  &  17$^h$47$^m$01.$^s$69 &   --34$^\circ$46'03.''0 &   0.$^d$42779  &   724.5659 &  15.$^m$38 &   1.$^m$34 &  0.$^m$21 &     EW \\
MM5-A V14  &  17$^h$47$^m$46.$^s$81 &   --34$^\circ$49'36.''7 &   2.$^d$09976  &   722.3363 &  15.$^m$49 &   1.$^m$18 &  0.$^m$28 &     EA \\
MM5-A V22  &  17$^h$47$^m$47.$^s$36 &   --34$^\circ$50'22.''3 &   0.$^d$35988  &   724.6137 &  15.$^m$63 &   1.$^m$27 &  0.$^m$38 &     EW \\
MM5-A V23  &  17$^h$47$^m$49.$^s$43 &   --34$^\circ$38'52.''7 &   0.$^d$40536  &   724.6236 &  15.$^m$67 &        --      &  0.$^m$29 &     EW \\
\end{tabular}
\end{center}
\pagebreak

\vspace{7pt}
\begin{center}
Table 6\\
Concluded\\
\vspace{7pt}

\begin{tabular}{lccrccccl}
Star ID  & $\alpha_{2000}$            & $\delta_{2000}$                &\multicolumn{1}{c}{$P$}       &T$_0$ --& $I$    & $(V{-}I)$ & $\Delta I$ & Type \\
OGLE      &                            &                                &           & 2~448~000 &        &        &       &               \\
MM5-A V89  &  17$^h$47$^m$50.$^s$04 &   --34$^\circ$45'07.''5 &   0.$^d$51219  &   724.4861 &  17.$^m$23 &   1.$^m$37 &  0.$^m$20 &     EW \\
MM5-A V90  &  17$^h$47$^m$08.$^s$17 &   --34$^\circ$43'16.''6 &   2.$^d$39688  &   724.0291 &  17.$^m$31 &   1.$^m$76 &  0.$^m$44 &     EA \\
MM5-A V91  &  17$^h$47$^m$25.$^s$95 &   --34$^\circ$45'24.''1 &   2.$^d$26790  &   722.6368 &  17.$^m$31 &   1.$^m$27 &  1.$^m$30 &     EA \\
MM5-A V92  &  17$^h$46$^m$50.$^s$16 &   --34$^\circ$45'55.''0 &   0.$^d$33091  &   724.6747 &  17.$^m$33 &   1.$^m$49 &  0.$^m$14 &     EW \\
MM5-A V94  &  17$^h$46$^m$54.$^s$60 &   --34$^\circ$39'45.''5 &   0.$^d$42404  &   724.6509 &  17.$^m$37 &   1.$^m$24 &  0.$^m$17 &     EW \\
MM5-A V95  &  17$^h$46$^m$50.$^s$14 &   --34$^\circ$44'54.''3 &   0.$^d$26055  &   724.6891 &  17.$^m$37 &   1.$^m$93 &  0.$^m$78 &     EW \\
\end{tabular}
\end{center}
\pagebreak
         
\vspace{7pt}
\begin{center}
Table 7\\
Eclipsing Variable Stars in the MM5-B field\\
\vspace{7pt}

\begin{tabular}{lccrccccl}
Star ID  & $\alpha_{2000}$            & $\delta_{2000}$                &\multicolumn{1}{c}{$P$}       &T$_0$ --& $I$    & $(V{-}I)$ & $\Delta I$ & Type \\
OGLE      &                            &                                &           & 2~448~000 &        &        &       &               \\
MM5-B V2   &  17$^h$47$^m$24.$^s$90 &   --35$^\circ$04'07.''7 &   0.$^d$29988  &   724.5883 &  14.$^m$84 &   1.$^m$26 &  0.$^m$81 &     EW \\
MM5-B V5   &  17$^h$47$^m$21.$^s$08 &   --35$^\circ$00'50.''5 &   0.$^d$44528  &   724.8067 &  15.$^m$03 &   1.$^m$17 &  0.$^m$31 &     EW \\
MM5-B V7   &  17$^h$47$^m$20.$^s$66 &   --34$^\circ$58'35.''3 &  16.$^d$48713  &   698.1640 &  15.$^m$20 &   1.$^m$82 &  0.$^m$22 &     EA \\
MM5-B V8   &  17$^h$47$^m$18.$^s$99 &   --34$^\circ$53'44.''8 &   0.$^d$35998  &   724.8240 &  15.$^m$27 &   1.$^m$24 &  0.$^m$06 &     EW \\
MM5-B V13  &  17$^h$46$^m$52.$^s$82 &   --34$^\circ$54'13.''7 &   0.$^d$39692  &   724.4552 &  15.$^m$56 &   1.$^m$36 &  0.$^m$43 &     EW \\
MM5-B V14  &  17$^h$47$^m$02.$^s$16 &   --35$^\circ$00'38.''3 &   1.$^d$24016  &   723.1572 &  15.$^m$58 &   1.$^m$15 &  0.$^m$77 &     EA \\
\end{tabular}
\end{center}
\pagebreak

\vspace{7pt}
\begin{center}
Table 7\\
Concluded\\
\vspace{7pt}

\begin{tabular}{lccrccccl}
Star ID  & $\alpha_{2000}$            & $\delta_{2000}$                &\multicolumn{1}{c}{$P$}       &T$_0$ --& $I$    & $(V{-}I)$ & $\Delta I$ & Type \\
OGLE      &                            &                                &           & 2~448~000 &        &        &       &               \\
MM5-B V92  &  17$^h$47$^m$46.$^s$65 &   --34$^\circ$55'22.''4 &   0.$^d$49450  &   724.5671 &  17.$^m$12 &   1.$^m$24 &  0.$^m$25 &     EW \\
MM5-B V93  &  17$^h$46$^m$57.$^s$06 &   --35$^\circ$00'50.''3 &   0.$^d$97253  &   724.0992 &  17.$^m$13 &   1.$^m$46 &  0.$^m$27 &     EW \\
MM5-B V94  &  17$^h$47$^m$40.$^s$85 &   --34$^\circ$55'40.''2 &   0.$^d$61598  &   724.0642 &  17.$^m$14 &   1.$^m$16 &  0.$^m$25 &     EW \\
MM5-B V95  &  17$^h$46$^m$53.$^s$25 &   --34$^\circ$59'09.''0 &   0.$^d$62983  &   724.3291 &  17.$^m$17 &        --      &  0.$^m$21 &    EW? \\
MM5-B V96  &  17$^h$47$^m$31.$^s$56 &   --35$^\circ$02'55.''5 &   0.$^d$80081  &   724.5985 &  17.$^m$16 &   1.$^m$23 &  0.$^m$29 &     EW \\
MM5-B V98  &  17$^h$47$^m$47.$^s$55 &   --35$^\circ$00'27.''6 &   0.$^d$70152  &   724.4071 &  17.$^m$17 &   1.$^m$21 &  0.$^m$50 &     EW \\
\end{tabular}
\end{center}

\pagebreak
         
\vspace{7pt}
\begin{center}
Table 8\\
Eclipsing Variable Stars in the MM7-A field\\
\vspace{7pt}

\begin{tabular}{lccrccccl}
Star ID  & $\alpha_{2000}$            & $\delta_{2000}$                &\multicolumn{1}{c}{$P$}       &T$_0$ --& $I$    & $(V{-}I)$ & $\Delta I$ & Type \\
OGLE      &                            &                                &           & 2~448~000 &        &        &       &               \\
MM7-A V6   &  18$^h$10$^m$48.$^s$74 &   --26$^\circ$01'20.''8 &   0.$^d$40395  &   724.6136 &  15.$^m$04 &   1.$^m$37 &  0.$^m$18 &     EW \\
MM7-A V9   &  18$^h$11$^m$14.$^s$23 &   --25$^\circ$51'54.''3 &   1.$^d$06942  &   723.2370 &  15.$^m$26 &   1.$^m$18 &  0.$^m$12 &     EW \\
MM7-A V10  &  18$^h$11$^m$05.$^s$56 &   --25$^\circ$57'46.''8 &   1.$^d$88386  &   722.8874 &  15.$^m$26 &   1.$^m$23 &  0.$^m$73 &     EA \\
MM7-A V12  &  18$^h$10$^m$22.$^s$38 &   --25$^\circ$55'29.''4 &   1.$^d$13987  &   724.7339 &  15.$^m$41 &   1.$^m$14 &  0.$^m$10 &     EW \\
MM7-A V13  &  18$^h$10$^m$48.$^s$04 &   --25$^\circ$53'08.''8 &   0.$^d$33346  &   724.7063 &  15.$^m$41 &   1.$^m$14 &  0.$^m$41 &     EW \\
MM7-A V14  &  18$^h$10$^m$56.$^s$61 &   --25$^\circ$49'11.''8 &   0.$^d$61198  &   724.4766 &  15.$^m$41 &   1.$^m$49 &  0.$^m$41 &     EW \\
\end{tabular}
\end{center}
\pagebreak

\vspace{7pt}
\begin{center}
Table 8\\
Continued\\
\vspace{7pt}

\begin{tabular}{lccrccccl}
Star ID  & $\alpha_{2000}$            & $\delta_{2000}$                &\multicolumn{1}{c}{$P$}       &T$_0$ --& $I$    & $(V{-}I)$ & $\Delta I$ & Type \\
OGLE      &                            &                                &           & 2~448~000 &        &        &       &               \\
MM7-A V105 &  18$^h$11$^m$16.$^s$01 &   --25$^\circ$59'48.''0 &   0.$^d$55982  &   724.9456 &  17.$^m$05 &   1.$^m$44 &  0.$^m$12 &     EW \\
MM7-A V106 &  18$^h$10$^m$55.$^s$29 &   --25$^\circ$49'56.''3 &   0.$^d$88103  &   723.9512 &  17.$^m$06 &   1.$^m$40 &  0.$^m$54 &     EW \\
MM7-A V107 &  18$^h$10$^m$35.$^s$56 &   --26$^\circ$00'47.''1 &   0.$^d$41950  &   724.8322 &  17.$^m$08 &   1.$^m$55 &  0.$^m$11 &     EW \\
MM7-A V108 &  18$^h$10$^m$50.$^s$42 &   --25$^\circ$53'07.''7 &   0.$^d$55069  &   724.6001 &  17.$^m$08 &   1.$^m$37 &  0.$^m$48 &     EW \\
MM7-A V110 &  18$^h$10$^m$37.$^s$43 &   --25$^\circ$52'51.''3 &   1.$^d$21045  &   723.4194 &  17.$^m$09 &   1.$^m$30 &  0.$^m$55 &     EA \\
MM7-A V111 &  18$^h$11$^m$17.$^s$21 &   --25$^\circ$47'46.''1 &   1.$^d$15640  &   723.7136 &  17.$^m$09 &   1.$^m$40 &  0.$^m$48 &     EA \\
\end{tabular}
\end{center}

\pagebreak

\vspace{7pt}
\begin{center}
Table 8\\
Concluded\\
\vspace{7pt}

\begin{tabular}{lccrccccl}
Star ID  & $\alpha_{2000}$            & $\delta_{2000}$                &\multicolumn{1}{c}{$P$}       &T$_0$ --& $I$    & $(V{-}I)$ & $\Delta I$ & Type \\
OGLE      &                            &                                &           & 2~448~000 &        &        &       &               \\
MM7-A V168 &  18$^h$10$^m$31.$^s$74 &   --25$^\circ$49'02.''2 &   0.$^d$43439  &   724.7797 &  17.$^m$79 &   1.$^m$40 &  0.$^m$22 &     EW \\
MM7-A V169 &  18$^h$10$^m$44.$^s$75 &   --26$^\circ$01'09.''2 &   0.$^d$45565  &   724.7095 &  17.$^m$79 &   1.$^m$35 &  0.$^m$47 &     EW \\
MM7-A V170 &  18$^h$10$^m$58.$^s$05 &   --25$^\circ$51'28.''8 &   0.$^d$50447  &   724.5688 &  17.$^m$79 &   1.$^m$51 &  0.$^m$41 &     EW \\
MM7-A V171 &  18$^h$11$^m$04.$^s$53 &   --26$^\circ$01'23.''2 &   0.$^d$41813  &   724.5779 &  17.$^m$81 &        --      &  0.$^m$14 &     EW \\
MM7-A V172 &  18$^h$10$^m$30.$^s$41 &   --25$^\circ$55'54.''3 &   0.$^d$50154  &   724.5558 &  17.$^m$81 &   1.$^m$36 &  0.$^m$13 &     EW \\
MM7-A V173 &  18$^h$10$^m$41.$^s$74 &   --25$^\circ$57'57.''4 &   1.$^d$03263  &   724.0017 &  17.$^m$82 &   1.$^m$64 &  0.$^m$28 &     EW \\
\end{tabular}
\end{center}

\vspace{7pt}
\begin{center}
Table 9\\
Eclipsing Variable Stars in the MM7-B field\\
\vspace{7pt}

\begin{tabular}{lccrccccl}
Star ID  & $\alpha_{2000}$            & $\delta_{2000}$                &\multicolumn{1}{c}{$P$}       &T$_0$ --& $I$    & $(V{-}I)$ & $\Delta I$ & Type \\
OGLE      &                            &                                &           & 2~448~000 &        &        &       &               \\
MM7-B V3   &  18$^h$12$^m$00.$^s$63 &   --26$^\circ$01'39.''5 &   0.$^d$72510  &   724.0914 &  15.$^m$01 &   1.$^m$10 &  0.$^m$31 &     EW \\
MM7-B V9   &  18$^h$11$^m$47.$^s$97 &   --25$^\circ$53'01.''6 &   0.$^d$88827  &   724.3877 &  15.$^m$52 &   1.$^m$20 &  0.$^m$07 &     EW \\
MM7-B V11  &  18$^h$11$^m$45.$^s$48 &   --25$^\circ$52'28.''6 &   0.$^d$61894  &   724.1685 &  15.$^m$57 &   1.$^m$21 &  0.$^m$06 &     EW \\
MM7-B V16  &  18$^h$11$^m$59.$^s$32 &   --25$^\circ$50'11.''1 &   0.$^d$47781  &   724.5812 &  15.$^m$70 &   1.$^m$15 &  0.$^m$05 &     EW \\
MM7-B V18  &  18$^h$11$^m$58.$^s$07 &   --25$^\circ$56'56.''4 &   0.$^d$34351  &   724.5979 &  15.$^m$77 &   1.$^m$22 &  0.$^m$17 &     EW \\
MM7-B V19  &  18$^h$11$^m$18.$^s$06 &   --26$^\circ$02'04.''2 &   0.$^d$53559  &   724.2816 &  15.$^m$82 &   1.$^m$08 &  0.$^m$16 &     EW \\
\end{tabular}
\end{center}
\pagebreak

\vspace{7pt}
\begin{center}
Table 9\\
Concluded\\
\vspace{7pt}

\begin{tabular}{lccrccccl}
Star ID  & $\alpha_{2000}$            & $\delta_{2000}$                &\multicolumn{1}{c}{$P$}       &T$_0$ --& $I$    & $(V{-}I)$ & $\Delta I$ & Type \\
OGLE      &                            &                                &           & 2~448~000 &        &        &       &               \\
MM7-B V56  &  18$^h$11$^m$28.$^s$31 &   --25$^\circ$49'30.''4 &   0.$^d$88241  &   724.4533 &  16.$^m$85 &   1.$^m$53 &  0.$^m$20 &     EW \\
MM7-B V58  &  18$^h$11$^m$56.$^s$52 &   --25$^\circ$59'01.''3 &   1.$^d$00021  &   723.4409 &  16.$^m$92 &   1.$^m$60 &  0.$^m$30 &      E \\
MM7-B V59  &  18$^h$12$^m$06.$^s$10 &   --25$^\circ$56'06.''9 &   0.$^d$38182  &   724.4553 &  16.$^m$92 &   1.$^m$60 &  0.$^m$44 &     EW \\
MM7-B V60  &  18$^h$12$^m$05.$^s$87 &   --25$^\circ$53'07.''5 &   4.$^d$33096  &   722.6181 &  16.$^m$94 &   1.$^m$75 &  0.$^m$55 &     EA \\
MM7-B V61  &  18$^h$11$^m$46.$^s$18 &   --26$^\circ$00'56.''1 &   1.$^d$77651  &   724.3151 &  16.$^m$98 &   1.$^m$46 &  0.$^m$19 &      E \\
MM7-B V62  &  18$^h$11$^m$30.$^s$20 &   --25$^\circ$49'18.''8 &   0.$^d$47980  &   724.7686 &  16.$^m$99 &   1.$^m$61 &  0.$^m$23 &     EW \\
\end{tabular}
\end{center}
\pagebreak

\vspace{7pt}
\begin{center}
Table 10\\
Miscellaneous Variable Stars in the MM5-A field\\
\vspace{7pt}

\begin{tabular}{lccrccccl}
Star ID  & $\alpha_{2000}$            & $\delta_{2000}$                &\multicolumn{1}{c}{$P$}       &T$_0$ --& $I$    & $(V{-}I)$ & $\Delta I$ & Type \\
OGLE      &                            &                                &           & 2~448~000 &        &        &       &               \\
MM5-A V1   &  17$^h$47$^m$57.$^s$55 &   --34$^\circ$47'34.''6 &  15.$^d$34369  &   720.1868 &  14.$^m$48 &        --      &  0.$^m$20 &      MISC \\
MM5-A V2   &  17$^h$47$^m$26.$^s$43 &   --34$^\circ$46'08.''4 &  19.$^d$31698  &   710.9421 &  14.$^m$64 &   2.$^m$13 &  0.$^m$10 &      MISC \\
MM5-A V4   &  17$^h$47$^m$45.$^s$72 &   --34$^\circ$39'43.''9 &  16.$^d$80174  &   715.5693 &  14.$^m$92 &   2.$^m$10 &  0.$^m$09 &      MISC \\
MM5-A V5   &  17$^h$47$^m$19.$^s$74 &   --34$^\circ$51'42.''0 &  58.$^d$24145  &   692.6656 &  15.$^m$02 &   1.$^m$88 &  0.$^m$11 &     MISC? \\
MM5-A V8   &  17$^h$46$^m$57.$^s$47 &   --34$^\circ$43'26.''1 &  35.$^d$31804  &   682.4152 &  15.$^m$19 &   2.$^m$10 &  0.$^m$06 &      MISC \\
MM5-A V10  &  17$^h$47$^m$19.$^s$51 &   --34$^\circ$42'39.''9 &   8.$^d$77242  &   714.8031 &  15.$^m$29 &   1.$^m$92 &  0.$^m$11 &      MISC Ell\\
\end{tabular}
\end{center}
\pagebreak
         
\vspace{7pt}
\begin{center}
Table 11\\
Miscellaneous Variable Stars in the MM5-B field\\
\vspace{7pt}

\begin{tabular}{lccrccccl}
Star ID  & $\alpha_{2000}$            & $\delta_{2000}$                &\multicolumn{1}{c}{$P$}       &T$_0$ --& $I$    & $(V{-}I)$ & $\Delta I$ & Type \\
OGLE      &                            &                                &           & 2~448~000 &        &        &       &               \\
MM5-B V1   &  17$^h$47$^m$02.$^s$98 &   --35$^\circ$02'17.''1 &  79.$^d$92082  &   687.4789 &  13.$^m$59 &   2.$^m$93 &  0.$^m$07 &      MISC \\
MM5-B V3   &  17$^h$47$^m$04.$^s$77 &   --35$^\circ$03'42.''4 &   5.$^d$97944  &   717.6404 &  14.$^m$86 &   1.$^m$80 &  0.$^m$11 &      MISC \\
MM5-B V6   &  17$^h$47$^m$37.$^s$01 &   --34$^\circ$57'05.''1 &  33.$^d$46190  &   674.9934 &  15.$^m$01 &   1.$^m$83 &  0.$^m$07 &      MISC \\
MM5-B V9   &  17$^h$47$^m$54.$^s$69 &   --34$^\circ$58'13.''7 &   7.$^d$73015  &   711.9391 &  15.$^m$31 &   1.$^m$75 &  0.$^m$09 &      MISC \\
MM5-B V11  &  17$^h$47$^m$05.$^s$80 &   --35$^\circ$00'05.''3 &  36.$^d$78492  &   672.5019 &  15.$^m$49 &   1.$^m$93 &  0.$^m$10 &      MISC \\
MM5-B V12  &  17$^h$47$^m$41.$^s$89 &   --34$^\circ$54'49.''2 &  33.$^d$10852  &   672.5665 &  15.$^m$52 &   1.$^m$66 &  0.$^m$13 &      MISC \\
\end{tabular}
\end{center}
\pagebreak

\vspace{7pt}
\begin{center}
Table 12\\
Miscellaneous Variable Stars in the MM7-A field\\
\vspace{7pt}

\begin{tabular}{lccrccccl}
Star ID  & $\alpha_{2000}$            & $\delta_{2000}$                &\multicolumn{1}{c}{$P$}      
&T$_0$ --& $I$    & $(V{-}I)$ & $\Delta I$ & Type \\
OGLE      &                            &                                &           & 2~448~000
&        &        &       &               \\
MM7-A V1   &  18$^h$10$^m$48.$^s$94 &   --25$^\circ$47'12.''3 &  18.$^d$29324  &   697.5578 &  14.$^m$20 &   2.$^m$05 &  0.$^m$09 &      MISC \\
MM7-A V3   &  18$^h$10$^m$32.$^s$26 &   --25$^\circ$54'34.''5 &   2.$^d$62541  &   724.0224 &  14.$^m$81 &   1.$^m$47 &  0.$^m$08 &      MISC \\
MM7-A V4   &  18$^h$10$^m$21.$^s$25 &   --25$^\circ$55'59.''9 &  57.$^d$00208  &   642.6414 &  14.$^m$84 &   1.$^m$89 &  0.$^m$12 &      MISC \\
MM7-A V5   &  18$^h$11$^m$05.$^s$50 &   --26$^\circ$01'17.''6 &  51.$^d$79081  &   652.1499 &  14.$^m$89 &        --      &  0.$^m$06 &      MISC \\
MM7-A V7   &  18$^h$10$^m$27.$^s$96 &   --25$^\circ$49'26.''3 &  12.$^d$41680  &   714.8600 &  15.$^m$06 &   2.$^m$23 &  0.$^m$07 &      MISC Ell?\\
MM7-A V8   &  18$^h$10$^m$18.$^s$67 &   --25$^\circ$51'46.''8 &  37.$^d$23882  &   690.1098 &  15.$^m$12 &   2.$^m$15 &  0.$^m$19 &      MISC \\
\end{tabular}
\end{center}
\pagebreak

\vspace{7pt}
\begin{center}
Table 12\\
Concluded\\
\vspace{7pt}

\begin{tabular}{lccrccccl}
Star ID  & $\alpha_{2000}$            & $\delta_{2000}$               
&\multicolumn{1}{c}{$P$}
&T$_0$ --& $I$    & $(V{-}I)$ & $\Delta I$ & Type \\
OGLE      &                            &                               
&           & 2~448~000
&        &        &       &               \\
MM7-A V100 &  18$^h$10$^m$32.$^s$31 &   --25$^\circ$48'19.''7 &  19.$^d$91045  &   708.9116 &  16.$^m$98 &   2.$^m$15 &  0.$^m$09 &      MISC \\
MM7-A V109 &  18$^h$10$^m$24.$^s$64 &   --25$^\circ$48'07.''1 &  12.$^d$01375  &   720.1642 &  17.$^m$08 &   2.$^m$16 &  0.$^m$24 &      MISC \\
MM7-A V114 &  18$^h$10$^m$19.$^s$51 &   --25$^\circ$54'30.''8 &  11.$^d$95630  &   712.6553 &  17.$^m$14 &   1.$^m$92 &  0.$^m$29 &      MISC \\
MM7-A V115 &  18$^h$11$^m$20.$^s$37 &   --25$^\circ$50'35.''1 &  10.$^d$50578  &   723.4280 &  17.$^m$14 &   2.$^m$06 &  0.$^m$17 &      MISC \\
MM7-A V116 &  18$^h$11$^m$08.$^s$89 &   --25$^\circ$50'49.''3 &  62.$^d$00406  &   629.9613 &  17.$^m$17 &   2.$^m$01 &  0.$^m$11 &      MISC \\
MM7-A V119 &  18$^h$11$^m$16.$^s$57 &   --25$^\circ$58'48.''3 &   5.$^d$97073  &   722.0105 &  17.$^m$21 &   1.$^m$87 &  0.$^m$11 &      MISC \\
\end{tabular}
\end{center}

\vspace{8pt}
\begin{center}
Table 12\\
Miscellaneous Variable Stars in the MM7-B field\\
\vspace{7pt}

\begin{tabular}{lccrccccl}
Star ID   & $\alpha_{2000}$            & $\delta_{2000}$                &\multicolumn{1}{c}{$P$}       &T$_0$ --& $I$    & $(V{-}I)$ & $\Delta I$ & Type \\
OGLE       &                            &                                &           & 2~448~000 &        &        &       &               \\
MM7-B V1   &  18$^h$12$^m$00.$^s$47 &   --25$^\circ$52'01.''7 &  12.$^d$10997  &   709.4626 &  13.$^m$69 &   1.$^m$96 &  0.$^m$16 &      MISC \\
MM7-B V2   &  18$^h$11$^m$30.$^s$16 &   --25$^\circ$55'48.''4 &  62.$^d$85072  &   665.5263 &  14.$^m$72 &   2.$^m$13 &  0.$^m$05 &      MISC \\
MM7-B V4   &  18$^h$11$^m$50.$^s$60 &   --25$^\circ$57'52.''5 &  57.$^d$55825  &   650.5233 &  15.$^m$04 &   2.$^m$24 &  0.$^m$12 &      MISC \\
MM7-B V5   &  18$^h$11$^m$45.$^s$56 &   --26$^\circ$01'08.''6 &  14.$^d$47485  &   717.5038 &  15.$^m$30 &   2.$^m$05 &  0.$^m$08 &      MISC \\
MM7-B V10  &  18$^h$11$^m$49.$^s$24 &   --25$^\circ$56'44.''9 &  23.$^d$21652  &   706.2233 &  15.$^m$54 &   2.$^m$14 &  0.$^m$04 &      MISC \\
MM7-B V15  &  18$^h$11$^m$26.$^s$29 &   --25$^\circ$55'06.''8 &  39.$^d$36907  &   679.5495 &  15.$^m$67 &   1.$^m$90 &  0.$^m$06 &      MISC \\
\end{tabular}
\end{center}
}
\pagebreak

\vspace*{12pt}
REFERENCES
\vspace{0.5cm}

\noindent{Kholopov, P.N. et al.},~{1985},~{"General Catalogue of Variable 
Stars. The Fourth Edition", Nauka, Moscow}

\noindent{Lasker, B.M. et al.},~{1988},~{ApJS},~{68},~{1}

\noindent{Paczy\'nski, B.},~{1986},~{ApJ},~{304},~{1}

\noindent{Udalski, A., Szyma\'nski, M., Stanek, K.Z., Ka\l u\.zny, J.,
Kubiak, M., Mateo, M., Krzemi\'nski, W., Paczy\'nski, B., and Venkat, R.}
{1994a},~{Acta Astron.},~{44},~{165}

\noindent{Udalski, A., Kubiak, M.,  Ka\l u\.zny, J., Szyma\'nski, M., Mateo, 
M., and Krzemi\'nski, W.},~{1994b},~{Acta Astron.},~{44},~{317 (Paper~I)} 

\noindent{Udalski, A., Szyma\'nski, M., Ka\l u\.zny, J., Kubiak, M., Mateo, M., 
and Krzemi\'nski, W.},~{1995a},~{Acta Astron.},~{45},~{1 (Paper~II)} 

\noindent{Udalski, A., Olech, A., Szyma\'nski, M., Ka\l u\.zny, J., Kubiak, M., 
Mateo, M., and Krzemi\'nski, W.},~{1995b},~{Acta Astron.},~{45},~{433 (Paper~III)} 

\noindent{Udalski, A., Olech, A., Szyma\'nski, M., Ka\l u\.zny, J., Kubiak, M., 
Mateo, M., Krzemi\'nski, W. and Stanek K.Z.},~{1996},~{Acta Astron.},~{46},~{51 (Paper~IV)}

\end{document}